\documentclass[twocolumn, prb, showpacs]{revtex4-1}

\usepackage{graphicx}
\usepackage{amsmath}
\usepackage{amssymb}
\usepackage{xcolor}
\usepackage{bbm}
\usepackage{bbold}
\usepackage{mathrsfs}
\newcommand{\kb} {\mathbf{k}}
\newcommand{\kt}{{{\mathbf{\tilde k}}}}

\newcommand\gv{\mathbf{g}}
\newcommand\Vv{\mathbf{V}}
\newcommand\tv{\mathbf{t}}
\newcommand\Gv{\mathbf{G}}
\newcommand\Sigmav{\mathbf{\Sigma}}
\newcommand\Gammav{\mathbf{\Gamma}}

\newcommand\tbb{\mathbbm{t}}
\newcommand\Sigmabb{\mathbb{\Sigma}}

\newcommand\Gbb{\mathbbm{G}}
\newcommand\Obb{\mathbbm{O}}

\begin{document}

\title{Impurity-induced magnetic moments on the graphene-lattice Hubbard model:\\ an inhomogeneous cluster DMFT study}
\author{M. Charlebois$^{1}$, D. S\'en\'echal$^{1}$, A.-M. Gagnon$^{1}$, A.-M. S. Tremblay$^{1,2}$}
\affiliation{$^1$ D\'epartement de Physique and RQMP, Universit\'e de Sherbrooke, Sherbrooke,
QC, Canada\\
$^{2}$Canadian Institute for Advanced Research, Toronto, Ontario, Canada.}
\date{\today}
\begin{abstract}
Defect-induced magnetic moments are at the center of the research effort on spintronic applications of graphene. Here we study the problem of a nonmagnetic impurity in graphene with a new theoretical method, inhomogeneous cluster dynamical mean field theory (I-CDMFT), which takes into account interaction-induced short-range correlations while allowing long-range inhomogeneities.
The system is described by a Hubbard model on the honeycomb lattice. The impurity is modeled by a local potential.
For a large enough potential, interactions induce local antiferromagnetic correlations around the impurity and a net total spin $\frac12$ appears, in agreement with Lieb's theorem.
Bound states caused by the impurity are visible in the local density of states (LDOS) and have their energies shifted by interactions in a spin-dependent way, leading to the antiferromagnetic correlations.
Our results take into account dynamical correlations; nevertheless they qualitatively agree with previous mean-field and density functional theory (DFT) studies.
Moreover, they provide a relation between impurity potential and on-site repulsion $U$ that could in principle be used to determine experimentally the value of $U$.
\end{abstract}

\maketitle

\section{Introduction}

The unique properties of graphene\cite{novoselov_electric_2004,castro_neto_electronic_2009} have been extensively studied during the last decade. They have important technological potential. 
Graphene is a good candidate material for spintronics because of its long-range room temperature spin transport,\cite{tombros_electronic_2007,yang_observation_2011,han_tunneling_2010}  its high carrier mobility and gate-tunability.\cite{castro_neto_electronic_2009,han_graphene_2014}

Defect-induced magnetic moments are at the center of this research effort on spintronics. 
Because of electron-electron interactions, local magnetism can emerge around zigzag edges,\cite{feldner_magnetism_2010,yazyev_magnetic_2008} nonmagnetic impurities and vacancies.\cite{kumazaki_nonmagnetic-defect-induced_2007,yazyev_defect-induced_2007} Different geometries of defects have been studied using mean-field and first-principle simulations.\cite{yazyev_emergence_2010} 
They have shown that different impurities should have (anti)ferromagnetic polarization if they are on the (opposite) same sublattice\cite{yazyev_magnetism_2008,yazyev_defect-induced_2007,santos_magnetism_2010} due to the Ruderman-Kittel-Kasuya-Yosida (RKKY) interaction.\cite{saremi_rkky_2007} 

In this paper we focus on the physics of isolated non-magnetic impurities. Localized states around such impurities have been predicted by analytic calculations\cite{pereira_disorder_2006,peres_electron_2007,peres_local_2009} and by density functional theory\cite{yazyev_defect-induced_2007} (DFT). The corresponding sharp peaks in the local density of states (LDOS) have been observed with scanning tunneling microscopy for vacancies on a graphite surface.\cite{ugeda_missing_2010}  In addition, non-magnetic impurities lead to magnetic correlations in their environment.\cite{kumazaki_nonmagnetic-defect-induced_2007,yazyev_defect-induced_2007} The net total spin of the atoms surrounding the impurity is $\frac12$, consistent with Lieb's theorem.\cite{lieb_two_1989} 
This net spin has been observed consistently in many numerical simulations,\cite{yazyev_emergence_2010,yazyev_defect-induced_2007,kumazaki_nonmagnetic-defect-induced_2007} and in experiments.\cite{nair_spin-half_2012}


In addition to mean-field studies of models,\cite{kumazaki_nonmagnetic-defect-induced_2007,yazyev_defect-induced_2007} DFT was used to simulate realistic impurities by taking into account lattice relaxation and the chemical nature of the impurity.\cite{santos_first-principles_2010,yazyev_defect-induced_2007,wehling_resonant_2010,wehling_adsorbates_2009,santos_universal_2012,santos_magnetism_2012,singh_magnetism_2009} However, DFT does not include dynamical correlations. To remedy this, a dynamical mean-field theory (DMFT)\cite{georges_hubbard_1992,jarrell_hubbard_1992,georges_dynamical_1996} calculation of the effect of a single impurity in graphene has been performed by Haase et al.\cite{haase_magnetic_2011} They found a ferromagnetic pattern around a single impurity that is quite different from previous mean-field or DFT calculations. These authors argue that an approach preserving correlations is necessary to account for Kondo physics, which cannot be simulated by mean-field theory. Here we test this hypothesis using an inhomogeneous extension of cluster DMFT (I-CDMFT) for the Hubbard model on the honeycomb lattice. In addition to using larger clusters, we relax the assumption made in Ref.~\onlinecite{haase_magnetic_2011} that the self-energy of the clusters surrounding the impurity cluster is the same as that of the system without the impurity. Instead, the self-energies are determined self-consistently and a vary large supercluster of 114 sites is used. 

This paper is organized as follows.
After we introduce the model in the following section,  we present in Sect.~\ref{sec:analytic} the $T$-matrix formalism used to calculate the effect of a single impurity in an infinite lattice in the noninteracting limit. 
This simple calculation will be useful to interpret I-CDMFT results. 
In Sect.~\ref{sec:cdmft}, the I-CDMFT technique is described in detail. 
The results of our simulations, presented in Sect.~\ref{results} and discussed in Sect.~\ref{discussion}, support the mean-field conclusions.

\section{Model}

\begin{figure}[h]
\begin{center}
\includegraphics[width=0.67\hsize]{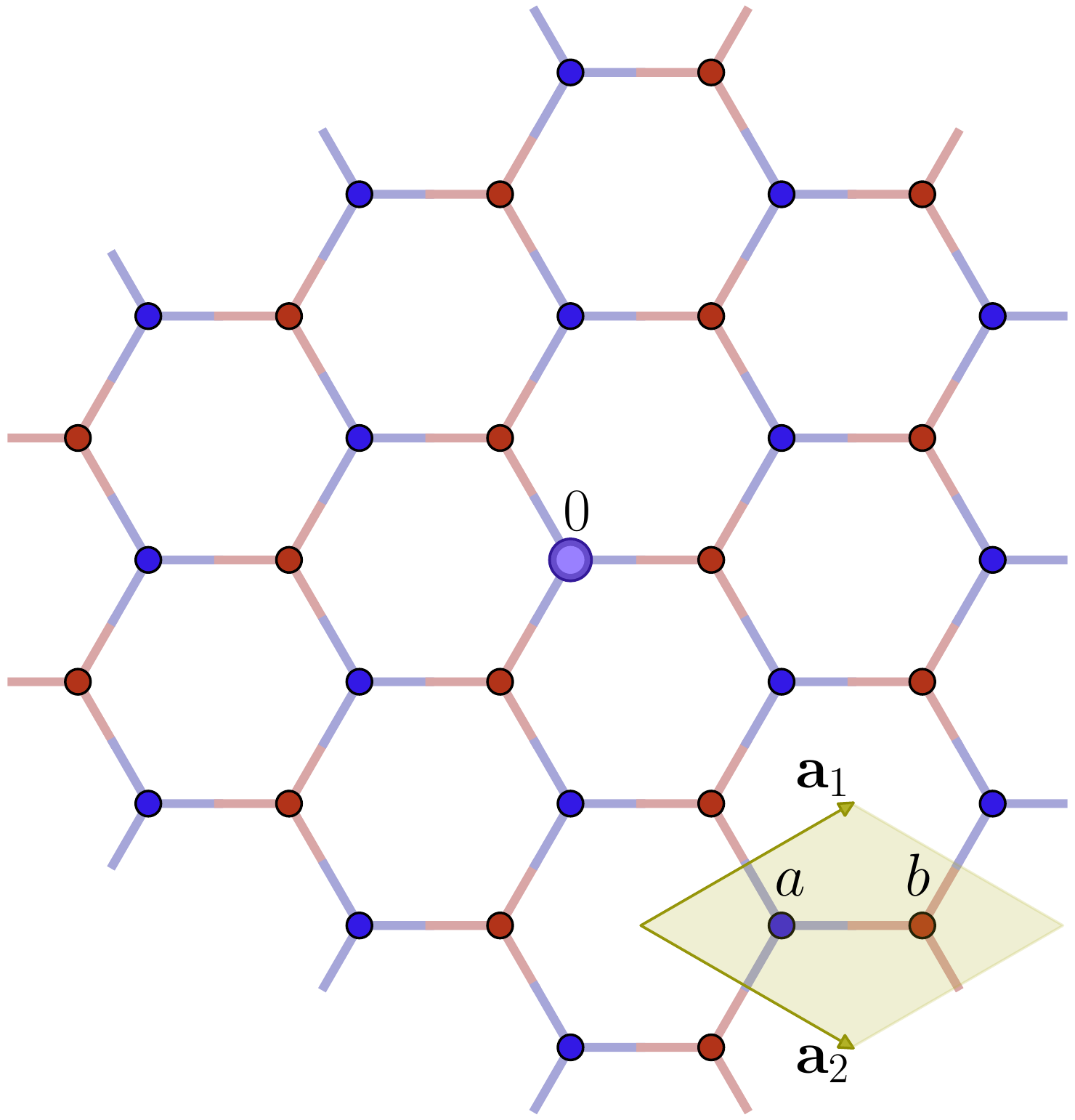}
\end{center}
\caption{(color online) The graphene lattice with an impurity at site $i=0$.
The two-site basis, with sublattice $a$ (blue) and $b$ (red), as well as the lattice vectors $\mathbf{a}_1$ and $\mathbf{a}_2$, are defined at the bottom-right corner.}
\label{fig:honey}
\end{figure}

To study the combined effect of a local impurity and of interactions in graphene, we study the Hubbard model with nearest-neighbor hopping $t$ and on-site repulsion $U$. The impurity is modeled as a local, spin-independent potential  $\epsilon_0$. The Hamiltonian is
\begin{eqnarray}\label{eq:Hubbard}
H =&-&t\sum_{\langle ij \rangle,\sigma}(c_{i,\sigma }^{\dag}c_{j,\sigma
}+\mathrm{H.c.})+U\sum_{i}n_{i\uparrow }n_{i\downarrow }  \notag \\ 
&-&\mu \sum_{i,\sigma} n_{i,\sigma } -\epsilon_0 \sum_{\sigma} n_{0,\sigma }
\end{eqnarray}%
as considered in Ref.~\citenum{haase_magnetic_2011}. 
$c_{i,\sigma }^{\dag}$ is the graphene $\pi$-band creation operator for an electron of spin $\sigma$ on site $i$; the index $i$ implicitly includes the lattice index $m$ and the sublattice index $\alpha=a,b$.
When $\lvert \epsilon_0 \rvert \gg t$, it is equivalent to an unrelaxed vacancy-type Hamiltonian, which has been studied in Refs.~\citenum{kumazaki_nonmagnetic-defect-induced_2007,yazyev_defect-induced_2007} using a mean-field approach.  We limit our simulations to half-filling, hence the chemical potential is $\mu=U/2$. All energy values are defined relative to the hopping amplitude $t=1$. Fig.~\ref{fig:honey} shows the graphene lattice with an impurity at site $i=0$. Note that this problem has $C_{3v}$ symmetry around the impurity. 
In this paper, we assume that the impurity resides on sublattice $a$.

\section{NONINTERACTING CASE}\label{sec:analytic}

Since the interesting physics appears in the limit $\epsilon_0 \gg U$, the $U=0$ limit is important to interpret the results. In this noninteracting case, a simple analytic solution to the problem is possible. The technique presented here is the common $T$-matrix formalism, which has been extensively used in graphene impurity problems.\cite{peres_local_2009,peres_electron_2007,peres_electronic_2006} 

In this section, we focus on the zero-temperature retarded Green function formalism. Since there are two inequivalent sites on the graphene lattice, it is convenient to write the Green function as a $2\times2$ matrix:
\begin{eqnarray}
  \gv_{mn}(\omega) =
  \left( {\begin{array}{cc}
   g^{aa}_{mn}(\omega) & g^{ab}_{mn}(\omega) \\
   g^{ba}_{mn}(\omega) & g^{bb}_{mn}(\omega) \\
  \end{array} } \right) 
\end{eqnarray}%
where $a$ and $b$ refer to the two different sublattices and $m$ and $n$ label the different unit cells. After setting $U=0$ in Hamiltonian (\ref{eq:Hubbard}), the impurity exact retarded Green function is given by
\begin{eqnarray}
\gv_{mn}(\omega) &=& \gv^0_{mn}(\omega) + \gv^0_{m0}(\omega)\Vv \gv_{0n}(\omega) \\
\mathrm{with~~}\Vv &\equiv& \begin{pmatrix}\epsilon_0 & 0 \\ 0 & 0\end{pmatrix}
\end{eqnarray}%
and where $\gv^0_{mn}(\omega)$ is the Green function without impurity. We use a lowercase $\gv$ to underline the noninteracting character of these Green functions. We can rewrite the above equation in terms of $\gv^0_{mn}(\omega)$ only:
\begin{equation}\label{GreenTMat}
\gv_{mn}(\omega) = \gv^0_{mn}(\omega) + \gv^0_{m0}(\omega) \textbf{T}_{00}(\omega) \gv^0_{0n}(\omega) 
\end{equation}%
where the $\textbf{T}$ matrix is given by
\begin{equation}
\textbf{T}_{00}(\omega) \equiv \Vv(1-\Vv \gv^0_{00}(\omega))^{-1}.
\end{equation}%
We know the Green function of pure graphene:
\begin{eqnarray}
\gv^0_{mn}(\omega) &=& \frac2N\sum_{\kb} e^{-i \kb\cdot(\textbf{r}_m-\textbf{r}_n)} \gv^0(\kb,\omega) \\
{\gv^0}(\kb,\omega) &=& 
\left( {\begin{array}{cc}
   z & t \phi_{\kb} \\
   t \phi^*_{\kb} & z \\
\end{array} } \right) ^{-1}
\label{G00} \\
\phi_{\kb} &=& 1 + e^{i \kb\cdot \textbf{a}_1} + e^{i \kb\cdot \textbf{a}_2} \\
z &\equiv& \omega + i \eta + \mu \label{z}.
\end{eqnarray}%
Here, $N$ is the (very large) number of atoms on the graphene sheet.
The three terms in $\phi_{\kb}$ correspond to the three nearest neighbors: the first is the hopping amplitude within the unit cell and the two exponentials refer to the hopping amplitudes to neighboring unit cells, separated by the basis vectors $\textbf{a}_1$ and $\textbf{a}_2$, respectively. In order to keep track of which site corresponds to which Green function, it is important to keep in mind that the choice of these two independent vectors dictates which basis we choose as our unit cell: it defines the mapping $i\to(m,\alpha)$.

The LDOS $A_{ii}(\omega)$ is the imaginary part of total Green function obtained from Eq. (\ref{GreenTMat}-\ref{G00}):
\begin{equation}\label{LDOS}
A_{ii}(\omega) = -\frac{1}{\pi}\operatorname{Im}g^{\alpha\alpha}_{mm}(\omega) 
= A^0_{ii}(\omega) + \delta A_{ii}(\omega)
\end{equation}%
where $A^0_{ii}(\omega)$, which comes from the first term of Eq.~\eqref{GreenTMat}, is the pure graphene density of states, and $\delta A_{ii}(\omega)$, coming from the second term of Eq.~\eqref{GreenTMat}, is the deviation caused by the impurity. 

As we increase $\epsilon_0$, the deviation $\delta A_{ii}(\omega)$ becomes more important. New poles come from the denominator of $\textbf{T}_{00}(\omega)$ around frequencies that satisfy $1=\epsilon_0 g^{0,aa}_{00}(\omega)$.
Since $A^0_{ii}(\omega)$ is small near $\omega=0$,  these new poles can create very sharp features (localized states) near that frequency.

\begin{figure}[h]
\begin{center}
\includegraphics[width=\hsize]{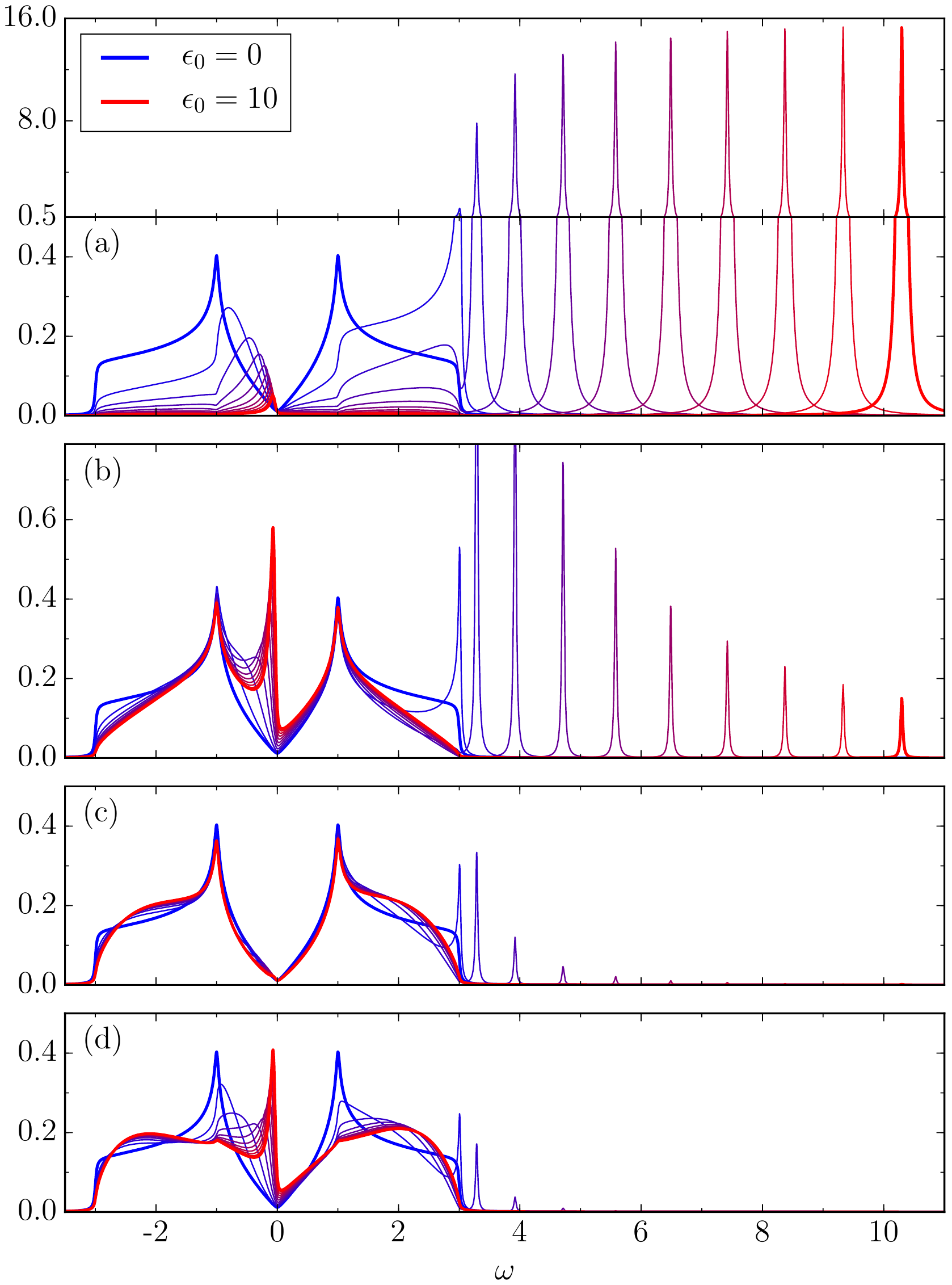}
\end{center}
\caption{(color online) (a) Evolution of the LDOS at the impurity site as a function of impurity potential $\epsilon_0$. Note that the $y$-axis scale is broken and has a larger scale on the upper panel to fit in the very large peaks. (b), the same, for a site next to the impurity. (c), the same, for a second neighbor, on the sublattice $a$. (d), the same, for a third neighbor, on sublattice $b$.}
\label{fig:evoluEps}
\end{figure}
Fig.~\ref{fig:evoluEps} shows the evolution of $A_{ii}(\omega)$ as a function of the impurity potential, from $\epsilon_0=0$ to $10$, for sites in the vicinity of the impurity. 
The only difference between the curve associated to $\epsilon_0=0$ and the analytic density of states of graphene from Hobson \& Nierenberg\cite{hobson_statistics_1953} comes from the Lorenzian broadening $\eta=0.02$ used here. For $\epsilon_0 \gg 1 $, a bound state appears slightly below the Fermi level ($\omega=0$) and an anti-bound state appears at frequency $\omega\sim\epsilon_0$. 
As $\epsilon_0\rightarrow\infty$, the LDOS at the impurity site vanishes in the range $-3 < \omega < 3$, just as for a vacancy. The bound state survives only on the sublattice on which there is no impurity. Particle-hole symmetry, present at $\epsilon_0 = 0$, is recovered at the limit $\epsilon_0\rightarrow \pm\infty$. Note that the LDOS for an attractive impurity is obtained by the substitution $\omega\rightarrow-\omega$.

\section{Inhomogeneous CDMFT}\label{sec:cdmft}

In the first subsection below, we recall the basics of CDMFT. Then we introduce the generalization of CDMFT to inhomogeneous systems that we introduce here. We finish by a short discussion of convergence issues. 

\subsection{CDMFT basics}

\begin{figure}[tbp]
\begin{center}
\includegraphics[width=0.9\hsize]{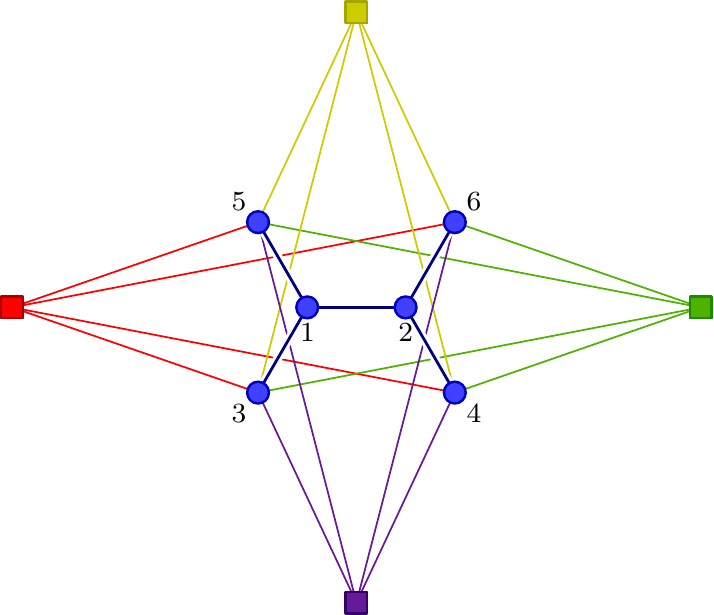}
\end{center}
\caption{(color online) Cluster-bath system used in this work. The numbered blue circles are the cluster sites per se.
The different bath-cluster hybridization terms are indicated by colored links.}
\label{fig:cluster}
\end{figure}

Let us first review the basics of CDMFT\cite{lichtenstein_antiferromagnetism_2000,kotliar_cellular_2001} for a homogeneous system.\cite{[{For reviews, see}],maier_quantum_2005,kotliar_electronic_2006,tremblay_pseudogap_2006,senechal__2012}
The method is based on a tiling of the infinite lattice by small, identical clusters, located at the sites of a superlattice.
In the exact diagonalization method (ED-CDMFT), the repeated cluster is coupled to a small set of noninteracting orbitals (the ``bath''), whose parameters are determined by a self-consistency condition that involves both the cluster's self-energy -- obtained here from the exact diagonalization technique -- and the infinite lattice, noninteracting Green function.
The cluster-bath system used in this work is illustrated on Fig.~\ref{fig:cluster}. One easily sees that the cluster can tile the graphene lattice (Fig.~\ref{fig:honey}). 

The cluster-bath system is described by the Anderson impurity model:
\begin{equation}\label{AIM}
\begin{split}
H_{\text{AIM}} = -\sum_{ij,\sigma}t_{ij}c_{i,\sigma }^{\dag}c_{j,\sigma
}+U\sum_{i}n_{i\uparrow }n_{i\downarrow }-\mu \sum_{i,\sigma
}n_{i,\sigma } \\
+\sum_{i\nu,\sigma} (\theta_{i \nu, \sigma} c_{i,\sigma }^{\dag}d_{\nu,\sigma} + \mathrm{H.c.}) + \sum_{\nu,\sigma} \varepsilon_{\nu, \sigma} d_{\nu,\sigma }^{\dag}d_{\nu,\sigma} 
\end{split}
\end{equation}
where $i,j$ denote the interacting sites of the cluster and $\nu$ the noninteracting bath sites. $\theta_{i \nu, \sigma}$ is the hopping amplitude from site $i$ to bath site $\nu$, and $\varepsilon_{\nu, \sigma}$ is the energy of bath site $\nu$. Note from Fig.~\ref{fig:cluster} that $\theta_{i \nu, \sigma}$ is zero for $i=1,2$ since there is only nearest-neighbor hopping in our model and bath sites are meant to represent the lattice environment of the cluster.

Our choice of cluster was guided by the possibility to surround the impurity site by 3 sites, in order to account for nearest-neighbor two-particle fluctuations, in a system small-enough to be solved repeatedly in a short time. Care must be taken in the parametrization of the bath sites in order to give this cluster enough degrees of freedom to represent the magnetic order we want to probe. 

The Green function of the cluster, extracted from the Anderson impurity model \eqref{eq:sigma} when traced over the bath sites, takes the following general form as a function of complex frequency $z$:
\begin{equation}\label{eq:sigma}
\Gv'^{-1}(z) = z - \tv - \Gammav(z) - \Sigmav(z)
\end{equation}
where the hybridization matrix $\Gammav(z)$ is
\begin{equation}\label{eq:hybridization}
\Gamma_{i j, \sigma}(z) = \sum_{\nu} \frac{\theta_{i \nu,\sigma}\theta_{j \nu,\sigma}^{*}}{z - \varepsilon_{\nu,\sigma}}.
\end{equation}%
We use a boldface matrix notation for the cluster Green function and related quantities: each cluster orbital (defined by site and spin) is associated with a row or column of the matrix.
In this work all matrices are diagonal in spin, since no spin-flip terms are present in $H$, even though the matrices associated with $\sigma=1$ and $-1$ will be different.
Using the same notation, the Green function on the infinite lattice, $\Gv(\kt,z)$, is specified by a frequency $z$ and a wavevector $\kt$ belonging to the Brillouin zone of the superlattice (the \textit{reduced} Brillouin zone):
The basic CDMFT approximation is to replace the full lattice self-energy by the computed, $\kt$-independent, cluster self-energy $\Sigmav(z)$. 
The lattice Green function is then approximated by
\begin{equation}\label{eq:latticeG}
\Gv^{-1}(\kt,z) = \Gv_0^{-1}(\kt,z) - \Sigmav(z),
\end{equation}
where $\Gv_0(\kt,z)=z-\tv(\kt)$ is the noninteracting Green function, but expressed in a mixed matrix/reduced wavevector form.\cite{maier_quantum_2005} 
In this form, $\tv(\kt)$ is composed of two different parts:
\begin{equation}\label{eq:t}
\tv(\kt) = \tv_c+\delta\tv(\kt),
\end{equation}
where $\tv_c$ is the hopping amplitude matrix within the cluster (it does not depend on $\kt$) and $\delta\tv(\kt)$ contains the hopping amplitudes between clusters.

The bath parameters are determined by a self-consistency condition stating that the cluster Green function $\Gv'(z)$ can also be obtained from the lattice Green function \eqref{eq:latticeG} by a Fourier transform:
\begin{equation}\label{eq:Gbar}
\Gv'(z) = \bar\Gv(z) \text{~~where~~} \bar\Gv(z) \equiv \frac{N_c}N \sum_\kt \Gv(\kt,z)
\end{equation}
($N$ is the total number of sites in the system and $N_c$ the number of sites on a cluster).
That condition cannot be satisfied exactly with the exact diagonalization technique, because the small number of bath orbitals does not provide enough parameters for the condition to be satisfied at all frequencies.
Instead, the bath parameters appearing in $\Gammav(z)$ are chosen so as to minimize the following distance function:
\begin{equation}\label{eq:distance}
d = \sum_{\substack{\mu,\nu \\ i\omega_n\leq i\omega_c}} \lvert (\Gv^{\prime-1}(i\omega_n)-\bar\Gv^{-1}(i\omega_n))_{\mu \nu} \rvert ^2.
\end{equation}
The sum is carried on the imaginary axis using Matsubara frequencies $\omega_n=2\pi n/\beta$ up to a cutoff $\omega_c=2$. We use a small ``fictitious'' temperature $\beta=50$ for this sum, even though the cluster Green function is computed numerically at zero temperature.
The variational parameters are all contained in the bath hybridization $\Gammav(\omega)$. 

In practice, the ED-CDMFT procedure follows this work flow:
\begin{enumerate}
\item Initial trial values of the bath parameters are chosen.
\item The ED is performed and the cluster Green functions $\Gv'(z)$ is computed, as well as the associated self-energy $\Sigmav(z)$, from Eq.~\eqref{eq:sigma}. The ED provides a representation of $\Gv'(z)$ for any complex frequency $z$.
\item The lattice Green function \eqref{eq:latticeG} is computed with the same self-energy as above, and it is then projected onto the cluster, giving $\bar\Gv$ defined in Eq.~\eqref{eq:Gbar}.
\item The bath parameters are updated by minimizing the distance function \eqref{eq:distance}.
In that expression $\bar\Gv$ is pre-computed, the self-energy $\Sigmav$ is considered fixed, whereas the hybridization function \eqref{eq:hybridization} is varied. Typically the Powell minimization technique is used at this step.
\item One goes back to step 2, until the bath parameters (or $\Gammav$) converge.
\end{enumerate}

Using ED-CDMFT with the cluster in Fig.~\ref{fig:cluster} (thus no impurity) we obtain an AFM transition at $U_c\cong2.8$. This is less than the value $U_c\cong3.9$ obtained from quantum Monte Carlo simulations.\cite{sorella_absence_2012} The accuracy in $U_c$ improves with cluster size, but we are limited in this respect because to the exponential growth of the Hilbert space with cluster size. Obtaining an accurate $U_c$ is not that important, as long as we stay in the range $U<U_c$, since we know that graphene is a semi-metal and not an antiferromagnet.

\subsection{Inhomogeneous CDMFT}

\begin{figure}[tbp]
\centering
\includegraphics[width=\hsize]{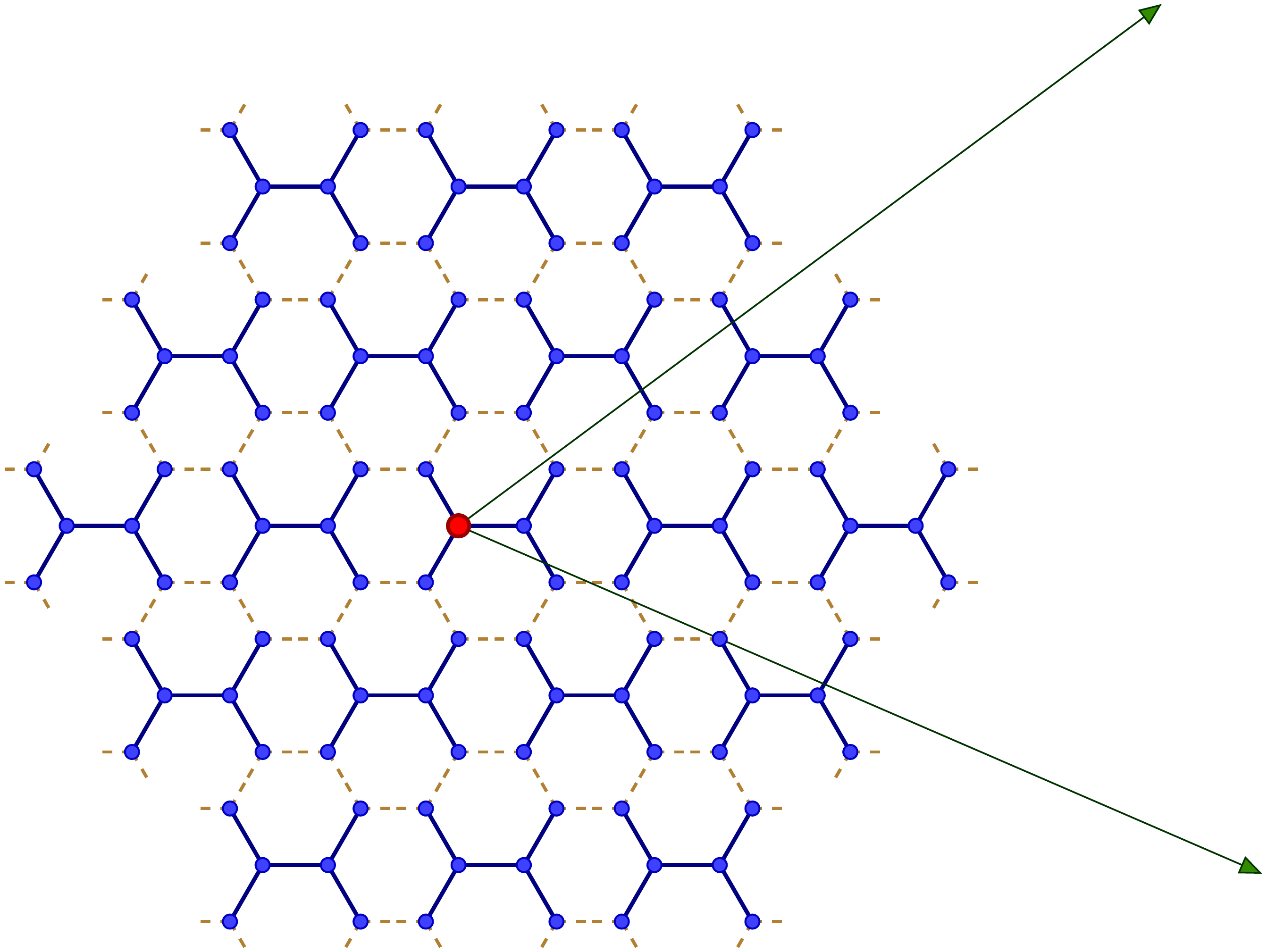}
\caption{(color online) The supercluster used in this work. The bath sites of each of the 19 6-site clusters are hidden for clarity.
The superlattice basis vectors are indicated with the green arrows. Inter-cluster links are the brown dotted lines.}
\label{fig:supercluster}
\end{figure}

From previous mean-field calculations, we know that antiferromagnetic correlations arising from the impurity go well beyond nearest neighbors.\cite{kumazaki_nonmagnetic-defect-induced_2007,yazyev_defect-induced_2007} 
In order to isolate the magnetism resulting from a single impurity, and at the same time avoid edge effects, we repeat the impurity
periodically, i.e., we define a superlattice with a large unit cell.
That unit cell is, however, too large for the exact diagonalization solver to manage.
We therefore assemble it from a number of smaller clusters; in other words, we define a large \textit{supercluster} composed of $19$ independent $6$-site clusters, as defined on Fig.~\ref{fig:supercluster}. We place one impurity on the middle cluster.
The supercluster, along with the impurity, is repeated according to the superlattice vectors shown on Fig.~\ref{fig:supercluster}.

We then use an inhomogeneous extension of CDMFT, similar to inhomogeneous DMFT,\cite{snoek_antiferromagnetic_2008} but using clusters instead of single sites in order to account for short-range, two-particle fluctuations.
We refer to this extension as I-CDMFT. Although we consider the specific case presented on Fig.~\ref{fig:supercluster}, keep in mind that the formalism here is very general and can be applied to any supercluster

The basic approximation of I-CDMFT is to replace the lattice self-energy matrix by a direct sum of the self-energy matrices of each independent cluster.
Since the $M$ clusters within the supercluster are different, we need to consider explicitly the Green function matrix for the supercluster, expressed as follows:
\begin{widetext}
\begin{eqnarray}\label{eq:G}
\Gbb^{-1}(\kt,z) =&& z-\tbb(\kt) - \Sigmabb(z) \label{totalGreen}\\ 
=&&\left( {\begin{array}{ccccc}
   z - \tv_{11}(\kt) - \Sigmav_1(z) & -\tv_{12}(\kt) & -\tv_{13}(\kt) & \ldots & 
   -\tv_{1M}(\kt)\\
   -\tv_{21}(\kt) & z - \tv_{22}(\kt) - \Sigmav_2(z) &  -\tv_{23}(\kt) & \ldots & 
   -\tv_{2M}(\kt)\\
   -\tv_{31}(\kt) & -\tv_{32}(\kt) & z - \tv_{33}(\kt) - \Sigmav_3(z) &  \ldots & 
   -\tv_{3M}(\kt)\\
   \vdots & \vdots & \vdots & \ddots & \vdots \\
   -\tv_{M1}(\kt) & -\tv_{M2}(\kt) & -\tv_{M3}(\kt) &  \ldots & z - \tv_{MM}(\kt) - \Sigmav_M(z)\\
\end{array} } \right) \nonumber
\end{eqnarray}%
\end{widetext}
where we use blackboard bold symbols for matrices having the dimensions of the supercluster.
Eq.~\eqref{eq:G} is analogous to Eq.~\eqref{eq:latticeG}, but here $\tbb(\kt)$ and $\Sigmabb(z)$ are the hopping and self-energy matrices for the supercluster, built from the cluster hopping matrices $\tv_{AB}(\kt)$ and self-energy matrices $\Sigmav_A(z)$ (the indices $A,B=1,\dots,M$ label the different clusters). 
Note that the matrix $\Sigmabb(z)$ is bloc diagonal. Each self-energy $\Sigmav_A(z)$ comes from the exact diagonalization procedure carried out independently on each cluster. On the other hand, $\tbb(\kt)$ is composed of two different parts:
\begin{equation}\label{eq:tt}
\tbb(\kt) = \tbb_\mathrm{sc} + \delta \tbb(\kt).
\end{equation}%
This is analogous to Eq.~\eqref{eq:t}, but $\kt$ is the reciprocal vector of the larger supercluster defined in Fig.~\ref{fig:supercluster}. $\tbb_\mathrm{sc}$ includes all intra-supercluster hopping terms and thus does not depend on $\kt$. Therefore $\delta \tbb(\kt)$ includes only inter-supercluster hopping terms. 

%


The superlattice Green function projected on a single supercluster is
\begin{equation}\label{latticeAvg}
\bar{\Gbb}(z) = \frac{N_c}N\sum_{\kt} \Gbb(\kt,z)
\end{equation}%
where $N_c=114$ is the number of sites on the supercluster.
The projection on a single cluster A is the $A^{th}$ diagonal block of $\bar{\Gbb}(z)$, noted $\bar{\Gv}_A(z)$.
This is the quantity that replaces the $\bar\Gv$ of ordinary CDMFT (Eq.~\ref{eq:Gbar}).
Like before, an approximate self-consistency is obtained by minimizing the distance function:
\begin{equation}
d = \sum_A\sum_{\substack{\mu \nu \\ i\omega_n\leq i\omega_N}} \lvert (\Gv_A'^{-1}(i\omega_n) - \bar\Gv_A^{-1}(i\omega_n))_{\mu \nu} \rvert ^2.
\label{distance}
\end{equation}
where $\Gv'_A$ is the Green function of cluster $A$ computed from Eq.~\eqref{eq:sigma}.
Note that the hybridization functions $\Gammav_A(z)$ -- and hence the bath parameters -- of the different clusters influence each other through the self-consistency condition because of the matrix inversions required to compute $\bar\Gbb$.

\subsection{Convergence issues}

To take into account both the impurity and interaction within the same simulation, we solve the complete supercluster in Fig.~\ref{fig:supercluster} using I-CDMFT. The variational parameters are the 32 hybridization terms ($\theta_{\nu\sigma}$) and the 8 bath energies ($\epsilon_{\nu \sigma}$) of Eq.~\eqref{AIM}, for each cluster. At each iteration of the CDMFT procedure, a total of $40\times19=760$ parameters is available to minimize the distance function~\eqref{distance}. 

Due to the large number of free parameters, care must be taken to avoid local (false) minima of the distance function. 
There is a small number of stable minima and one can easily identify which ones are not physical through symmetry considerations. It is recommended to first impose constraints between the $760$ variational parameters to obtain a first guess of the solution and to relax these constraints afterwards. In order to converge, it is necessary to start from a trial point not too far from the solution. A known CDMFT solution for a particular parameter set ($U$, $\epsilon_0$) in Eq.~\eqref{eq:Hubbard} constitutes a good guess for a slightly different parameter set. Using this approach, we can sweep the whole parameter space. At each step, we must fully converge the CDMFT procedure in order to obtain a solid starting solution for the next step. 

\section{I-CDMFT Results}
\label{results}

We first present the results for the semimetallic phase in the presence of interactions but without impurity. The more general case follows. 

\subsection{Without impurity ($\epsilon_0=0$)}
\label{U2e0}

\begin{figure}[tbp]
\begin{center}
\includegraphics[width=\hsize]{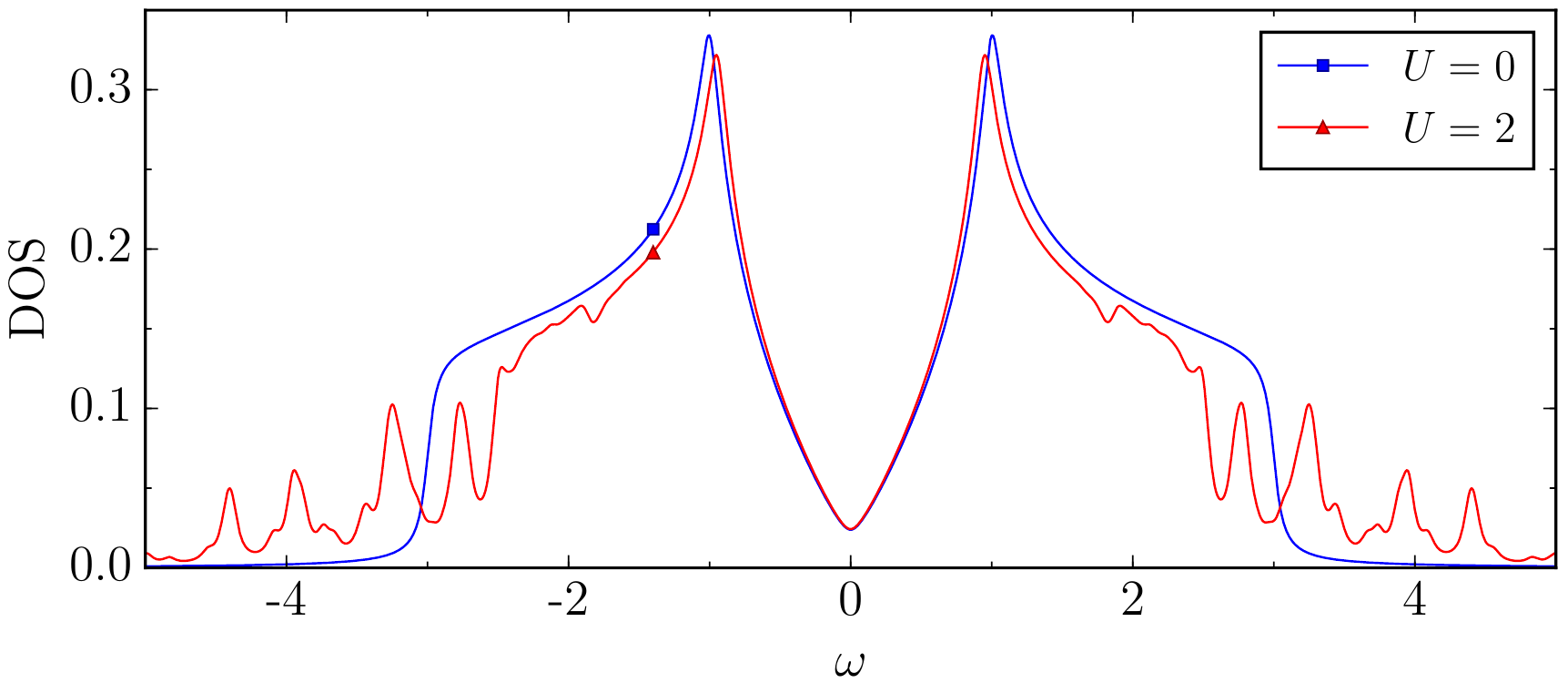}
\end{center}
\caption{(color online) Density of states for the noninteracting case (blue, exact) and $U=2$ (red, CDMFT) in the absence of impurity.
A Lorenzian broadening $\eta=0.05$ was used.}
\label{fig:U0_U2_eps0}
\end{figure}

The LDOS $A_{ii}(\omega)$ for $U=2$ and no impurity ($\epsilon_0=0$) is compared to the noninteracting solution ($U=0, \epsilon_0=0$) on Fig.~\ref{fig:U0_U2_eps0}. At low energies ($\lvert \omega \rvert \lesssim 1.5$), the LDOS from CDMFT is similar to the analytic LDOS without interaction. The van Hove singularities are shifted towards lower frequencies and have a smaller amplitude, but most of the low-energy physics is the same. At higher energies, there is a broadening of the band and sharp features appear. The former is to be expected, but the latter are an artefact of the finite-size of the cluster, a limitation of the exact diagonalization method. Indeed, the number of bath sites attached to the cluster is small, hence the energy spectrum is discrete, far from that of a continuous bath. The higher-energy physics is less accurate and we will therefore focus on the low-energy physics in the following discussion. Note that the interaction does not increase the Fermi velocity around the Fermi Level. An increase of the Fermi velocity is observed only when nearest-neighbor or longer range interactions are added.\cite{wu_phase_2014,wu_interacting_2010,gonzalez_marginal-fermi-liquid_1999,elias_dirac_2011} The shift of van Hove singularities is consistent with other DMFT studies of graphene.\cite{jafari_dynamical_2009}

\subsection{With impurity ($\epsilon_0\neq0$)}\label{U2e10}

\begin{figure}[tbp]
\includegraphics[width=\hsize]{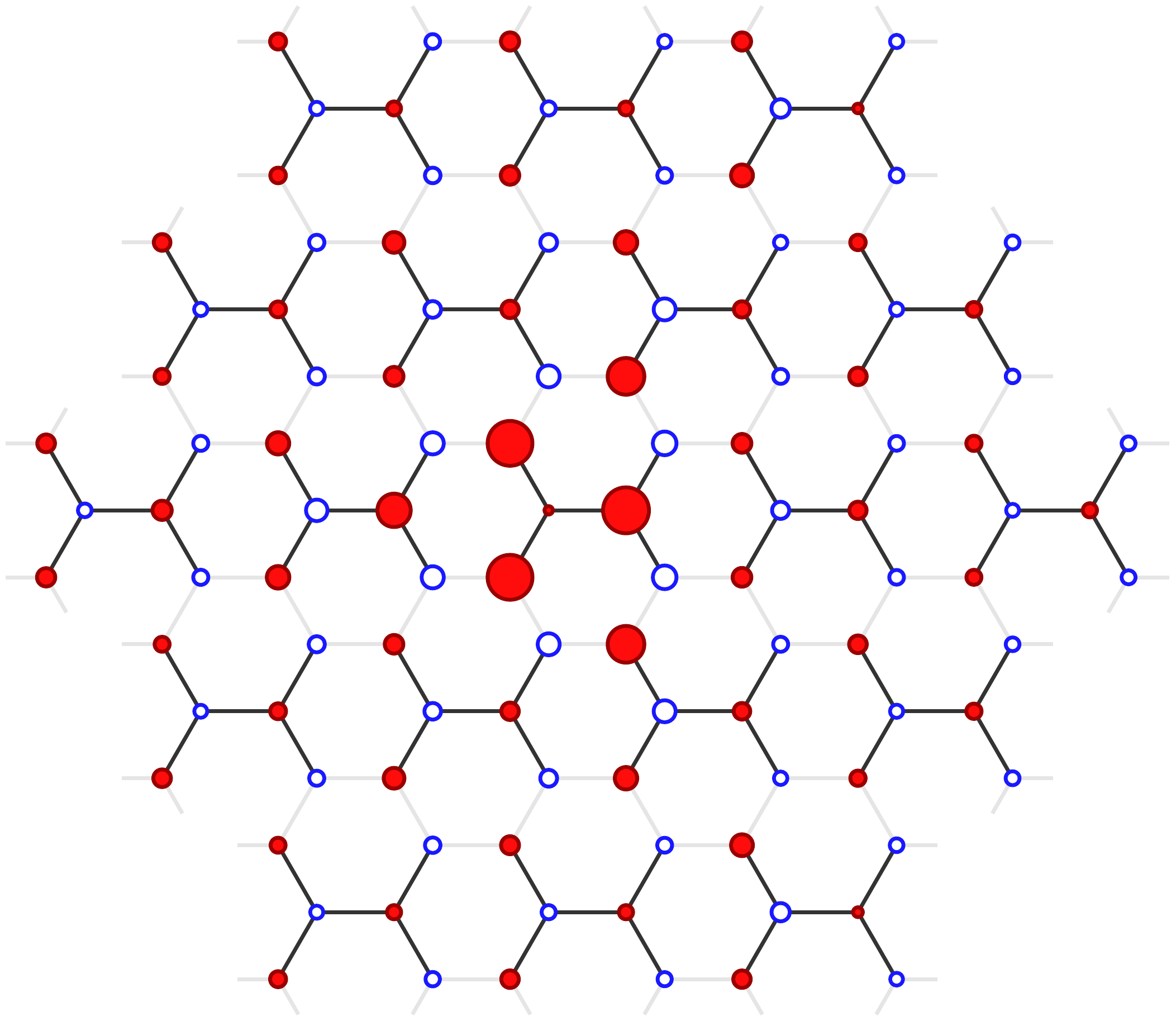}
\caption{(color online) Net spin of each site in the I-CDMFT solution for $U=2$ and $\epsilon_0=11$, represented by colored circles (blue is down, red is up). The area of each circle is proportional to $\langle S_i\rangle$, which ranges from $0.00315$ to $0.0771$.}
\label{fig:AFM_U2}
\end{figure}

We next consider the combined effect of interactions with a large impurity potential $\epsilon_0$. The I-CDMFT solution shows a local magnetization around the defect. The magnetization on site $i$ can be probed by this operator:
\begin{equation}
\hat{S}_{i} = \frac12\sum_{\sigma=\pm1} \sigma n_{i \sigma} 
\end{equation}%
and is represented on Fig.~\ref{fig:AFM_U2}. There is a local antiferromagnetic moment around the impurity site. The resulting antiferromagnetic pattern has been observed in other theoretical studies of the vacancy defect in graphene using mean-field\cite{kumazaki_nonmagnetic-defect-induced_2007,leong_effects_2014} or density functional theory.\citealp{yazyev_defect-induced_2007} The $120^\circ$ rotation symmetry is not exactly found in our solution since both the cluster and the supercluster do not have that symmetry. But the adjustable bath parameters manage to restore it almost completely.
Most of the features observed in the mean-field approach are found here. 

Sublattices $a$ and $b$ have opposite spin polarizations, except at the impurity site. Since $\epsilon_0$ is finite, this site is partially occupied. It has the same spin-polarization as its nearest neighbors.
The AFM order is enhanced because the impurity is periodically repeated. 

We can define operators $\hat{M}$ and $\hat{S}$ that measure the lattice-averaged AFM order parameter per site $\langle \hat{M} \rangle$ and the total spin $\langle \hat{S} \rangle$:
\begin{equation}
\hat{M} = \frac1{N_c}\left\{\sum_{i \in a}   \hat{S}_{i} -\sum_{i \in b}  \hat{S}_{i}\right\}, \qquad\qquad
\hat{S} =  \sum_{i} \hat{S}_{i}.
\end{equation}
Lattice averages of operators defined in second quantization $\hat{O}=\sum_{ij} O_{ij} c_i^\dag c_j$ are\cite{fetter_quantum_2003}
\begin{equation}\label{eq:averages}
\langle \hat O\rangle = \mathrm{Tr}[\Obb\Gbb]
\end{equation}
where $\Gbb$ is the Green function on the supercluster and the trace involves (1) a sum over supercluster sites, (2) an average over wavevectors in the reduced Brillouin zone and (3) an integral over frequencies, taken along the imaginary axis.\cite{senechal__2012,senechal_introduction_2008} 

\begin{figure}[tbp]
\includegraphics[width=\hsize]{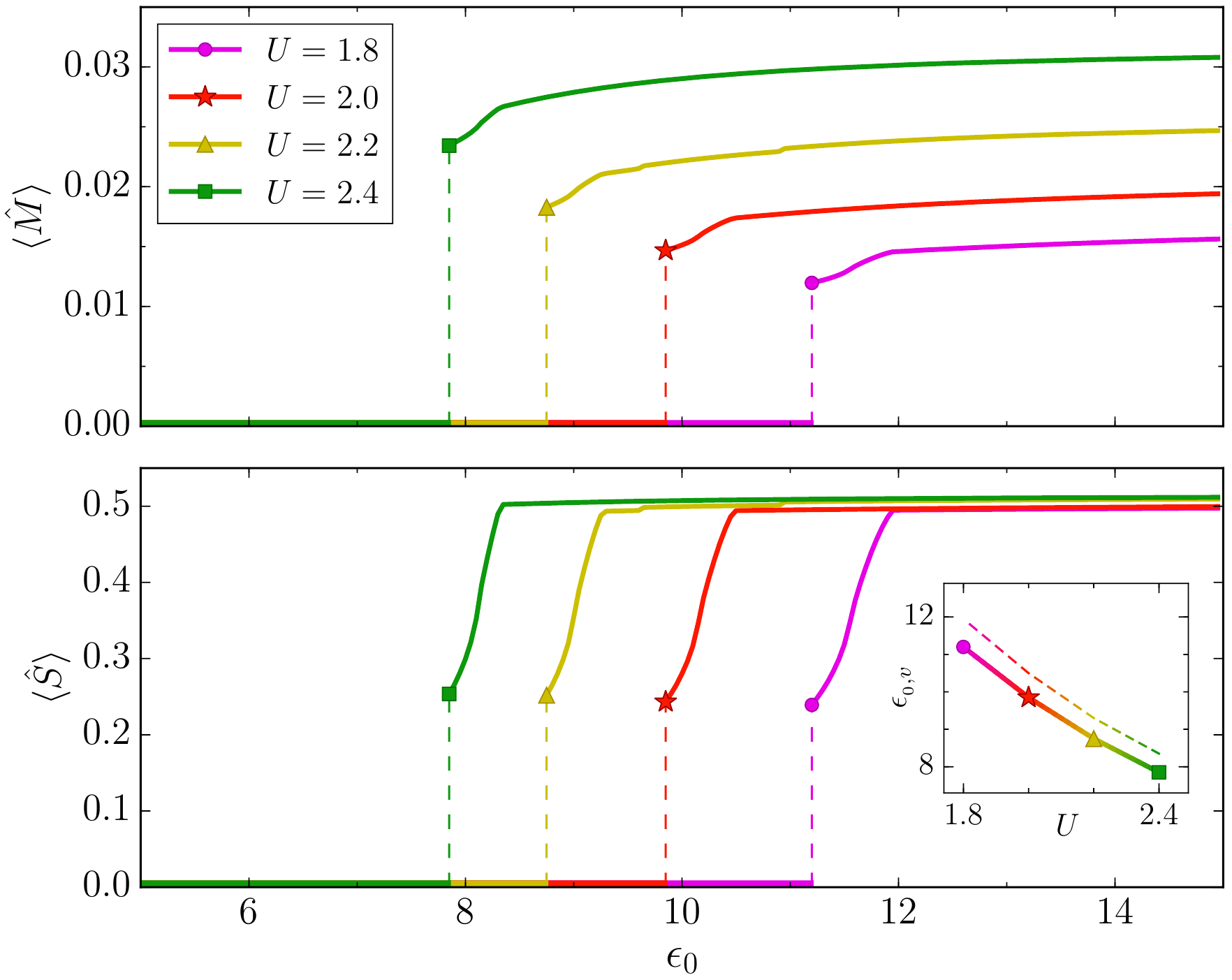}
\caption{(color online) Antiferromagnetic order parameter per site $\langle\hat M\rangle$ of the supercluster (top panel) and total spin $\langle\hat S\rangle$ (bottom panel), as a function of impurity potential $\epsilon_0$. The inset shows the relation between $U$ and the threshold values $\epsilon_{0,c}$ (full line) and $\epsilon_{0,v}$ (dashed line).}
\label{fig:AFM_nc}
\end{figure}

$\langle \hat{M} \rangle$ and $\langle \hat{S} \rangle$ are plotted on Fig.~\ref{fig:AFM_nc} as a function of the impurity potential $\epsilon_0$ for different $U$. For every $U$, there is no apparent order for $\epsilon_0$ lower than a critical value ($\epsilon_{0,c}$). At $\epsilon_{0,c}$, the net spin takes the value $\langle \hat{S} \rangle=\frac14$ and then grows stronger with $\epsilon_0$. For $\epsilon_0$ higher than a threshold value ($\epsilon_{0,v}$), the resulting magnetic state reaches the value $\langle \hat{S} \rangle = \frac12$ for the total spin. This is in perfect agreement with Lieb's theorem.\cite{lieb_two_1989} This theorem states that, on a bipartite lattice, the total spin of the ground state of the repulsive Hubbard model is $\frac12$, times the difference in the number of atoms of the two sublattices. Hence, the plateau in $\langle \hat{S} \rangle$ at $\epsilon_0\geqslant\epsilon_{0,v}$ is the signature of the magnetic state obtained for a vacancy defect. The kink in $\langle \hat{M} \rangle$ at $\epsilon_{0,v}$ is also a signature that the impurity then behaves as a vacancy. Both $\epsilon_{0,c}$ and $\epsilon_{0,v}$ are monotonic functions of $U$: knowing $\epsilon_{0,c}$ and $\epsilon_{0,v}$ defines a unique $U$, as shown in the inset of Fig.~\ref{fig:AFM_nc}.

\begin{figure*}
	\includegraphics[width=1\textwidth]{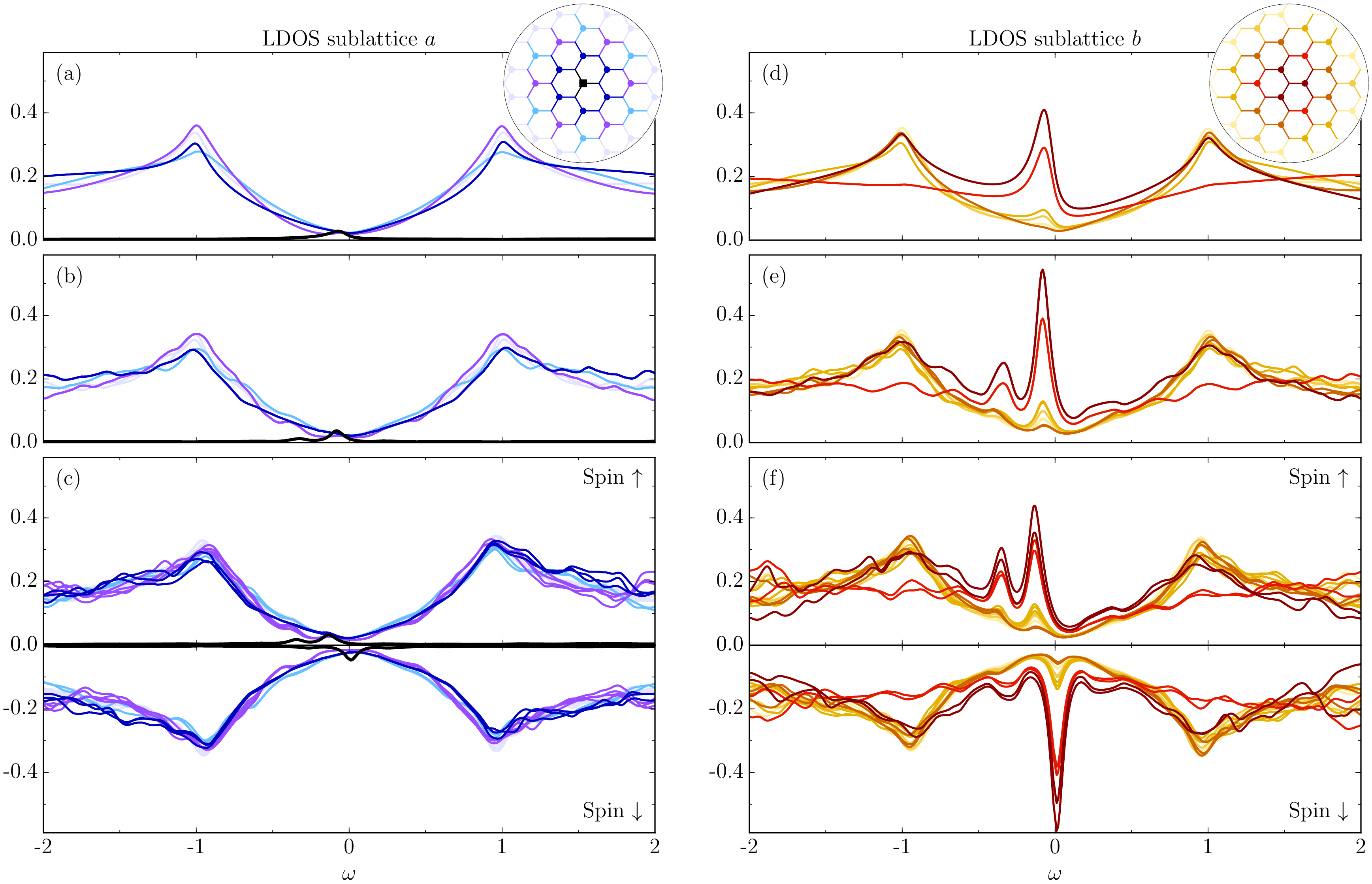}
	\caption{(color online) LDOS for sublattice $a$ (left panels) and for sublattice $b$ (right panels) for sites in the vicinity of the impurity. The impurity is on one of the sites of sublattice a. The distance from the impurity is color coded: the insets define the colors associated with each position around the impurity (black square). 
		The LDOS for sites closer to the impurity is darker. Panels (a) and (d): LDOS calculated from the $T$-matrix formalism presented in section~\ref{sec:analytic} with $\epsilon_c=11$. Panels (b) and (e): LDOS calculated with I-CDMFT, but without interactions ($U=0$, $\epsilon_c=11$). Panels (c) and (d): LDOS calculated with I-CDMFT and interactions ($U=2$, $\epsilon_c=11$) for both up and down spin as indicated on the figure. A Lorenzian broadening $\eta=0.05$ is used and the sign of the LDOS is reversed for spin down in the lower panels .}
	\label{fig:CDMFT_DOS}
\end{figure*}
\begin{figure}[h]
	\includegraphics[width=\hsize]{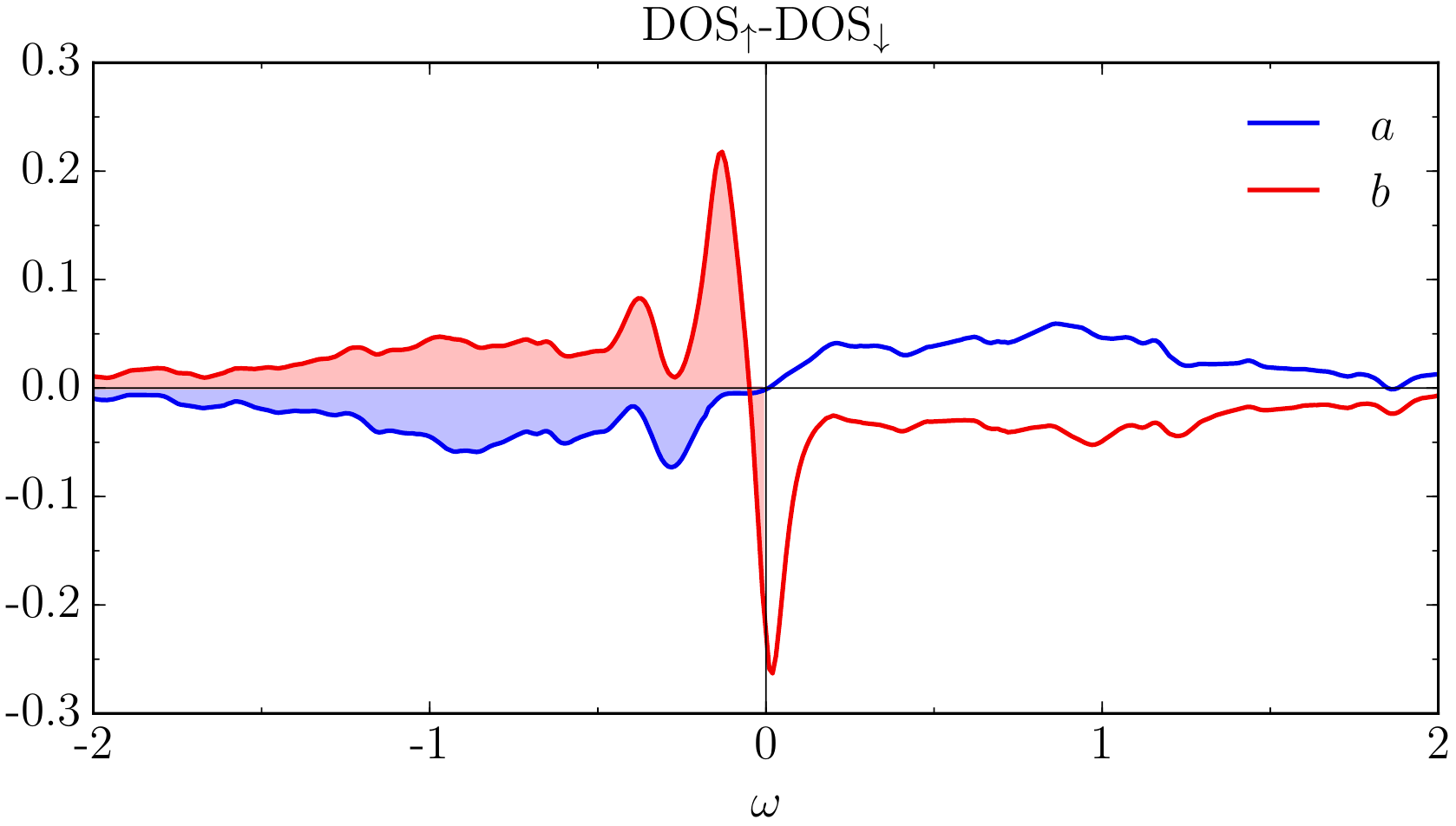}
	\caption{(color online) Difference between the density of states for spin-up and spin-down for sublattices $a$ (blue) and $b$ (red). Each curve is the sum of the LDOS over all sites of a given sublattice within the supercluster (57 sites for each sublattice).
		The shading represents the sum over frequencies up to the Fermi level and highlights the contribution of each sublattice to the spin-polarization.}
	\label{fig:diff}
\end{figure}

To gain insight into the physics, we present results for the LDOS. Fig.~\ref{fig:CDMFT_DOS} compares the LDOS for the noninteracting case (panels a and d at the top) with the I-CDMFT results (bottom panels), for the neighboring sites of the impurity. The middle panel data (b and e) can be obtained either from the $T$-matrix formalism with the impurity repeated with the superlattice vector of Fig.~\ref{fig:supercluster}, or from I-CDMFT if we set $U=0$. We show this LDOS to identify the effect of impurity repetition. The main effect of this repetition is to add oscillations and enhance the impurity amplitude.  The color of each LDOS curve is a function of the distance from the impurity. The insets show the relation between color and position. Without interaction, all the sites at the same distance of the impurity are equivalent and have the same LDOS, due to the $C_{3v}$ symmetry. This symmetry should remain even for finite $U$, but as seen on the bottom panels, the LDOS from different sites at the same distance (same color) of the impurity are slightly different. This is because the cluster and supercluster do not have $C_{3v}$ symmetry.

The bottom panels, c and f, of Fig.~\ref{fig:CDMFT_DOS} show that adding interactions spontaneously breaks time-reversal symmetry, as shown on Fig.~\ref{fig:AFM_U2}. The LDOS for spin down is plotted upside down to allow easier comparison with the spin up results. Because the spin-down bound state in Fig.~\ref{fig:CDMFT_DOS}f is shifted to higher frequencies, it eventually crosses the Fermi level ($\omega=0$) and becomes partly unoccupied, contrary to the spin-up bound state. This bound state spin imbalance is at the origin of the enhanced magnetism on sublattice $b$, i.e., the sublattice without impurity. 
Effectively, only sites on sublattice $b$ have this bound state and the concomitant enhanced spin-polarization. Since there is no bound state on sublattice $a$, the spin-polarization there comes entirely from the band distortion, which differs for spins up and down. This distortion in LDOS can be observed directly on the bottom panel of Fig.~\ref{fig:CDMFT_DOS}, but it is easier to observe if we plot the difference between spin up and spin down density of states. This LDOS difference, summed over the $57$ different sites of the same sublattice of our supercluster, is shown on Fig.~\ref{fig:diff} for each sublattice. The integrated area up to the Fermi level corresponds to the spin polarization of each sublattice. There is a contribution from band distortion on both sublattices. Note that the enhanced spin polarization resulting from the bound state on sublattice $b$ is reduced by the sum over the sites on this figure. For sites very close to the impurity, the bound state spin polarization dominates over the band distortion for this sublattice. The spin polarization on sublattice $b$ is therefore stronger  than on sublattice $a$, as observed on Fig.~\ref{fig:AFM_U2}.

An animated version of Fig.~\ref{fig:CDMFT_DOS} along with Fig.~\ref{fig:AFM_U2} and \ref{fig:AFM_nc} can be obtained online~\cite{__????}. Each frame of the animation corresponds to different values of $\epsilon_0$, which range from $0.0$ to $14.8$. The $760$ converged variational parameters, obtained from the I-CDMFT solution, used to generate this animation, can also be obtained online~\cite{__????-1}.


\section{Discussion} 
\label{discussion}

It has long been realized that impurities may help reveal underlying interactions in an otherwise featureless metallic state.\cite{kotov_electron-electron_2012} Our study confirms previous results that even if $U$ is not large enough to lead to an antiferromagnetic state, non-magnetic impurities can antiferromagnetically polarize their environment in the semi-metallic phase and  lead to a net spin $1/2$. Although the symmetry cannot be spontaneously broken for a single impurity, the antiferromagnetic correlations discussed in this paper can be revealed, for example, in NMR experiments because the small applied magnetic field selects a direction for the net spin.   

How impurities on one sublattice lead to antiferromagnetic correlations can be qualitatively understood as follows.  Antiferromagnetism naturally arises in graphene due to the perfect nesting of Dirac cones. As discussed in Ref.~\onlinecite{bercx_magnetic_2009,kotov_electron-electron_2012,kumazaki_nonmagnetic-defect-induced_2007}
the spin susceptibility diverges because at half-filling ($\mu=U/2$) there is perfect nesting for the wave vector $\mathbf{Q}=(0,0)$, corresponding to intra unit cell antiferromagnetic fluctuations, hence the Lindhard function is logarithmically divergent for that wave vector, namely
\begin{equation}
\chi^0_{\text{AFM}}(\mathbf{Q}) \sim \rho(0) \ln\left(\frac{W}{T}\right)
\label{lindhard}
\end{equation}
where $\rho(0)$ is the density of states at the Fermi level, $W$ is the bandwidth (here $W=6t$) and $T$ is the temperature. In an RPA expression, this result would imply that for a sufficiently low temperature, the susceptibility would diverge for arbitrarily small $U$. However, since $\rho(0)=0$ in pure graphene, one in fact needs a finite critical $U_c$ to observe antiferromagnetism. In our model, increasing the potential of the impurity $\epsilon_0$ brings the bound state closer to the Fermi level (Fig.~\ref{fig:evoluEps}). Hence, with the addition of an impurity, the density of state around the Fermi level becomes finite, thereby revealing the nesting instability. For a critical value of the impurity potential $\epsilon_{0}=\epsilon_{0,c}$, the RPA spin susceptibility 
diverges and antiferromagnetism appears. Since for a larger $U$, the RPA susceptibility diverges for a smaller value of $\rho(0)$, the threshold value of $\epsilon_{0}$ decreases with increasing $U$, consistent with Fig.~\ref{fig:AFM_nc}.

While the environment of the impurity is antiferromagnetically correlated, one clearly sees from Fig.~\ref{fig:AFM_U2} that when the local potential $\epsilon_0$ is finite, in other words when the impurity site not completely empty, the impurity site itself is ferromagnetically correlated with its neighbors. This agrees with the DMFT results of Ref.~\citenum{haase_magnetic_2011} that we set out to verify, but only for the nearest-neighbors.  In that paper,\cite{haase_magnetic_2011} they considered a cluster containing the impurity site and only one of its nearest neighbor that they attached to an environment whose baths were those of the infinite system. The LDOS on the impurity (present on sublattice $a$) contains the bound state responsible for the enhanced spin polarization of sublattice $b$. Since the impurity site is the only site on sublattice $a$ where this bound state is strong enough to dominate the spin polarization, it is the only site of sublattice $a$ that has the same spin polarization as sublattice $b$. The residual resonance on the impurity site is then responsible for the ferromagnetic correlation of the impurity and its nearest neighbors. For every other site, sublattices a and b have opposite spin polarizations. This is not captured by the approach of Ref.~\citenum{haase_magnetic_2011} that does not allow the bath on clusters beyond nearest neighbors to adjust to the presence of the impurity cluster. Our choice of cluster within the supercluster was motivated by the desire to have the impurity connect to all its nearest neighbors directly on the cluster and not only through the self-consistency relation. The need for the supercluster itself is clear when one notices that it is necessary to leave to the $a-b$ spin polarization on the impurity-containing cluster the freedom to be different from the $a-b$ spin polarization on other clusters.

It is important to impose periodic boundary conditions on the supercluster even if this means periodically repeating the impurity. Some tests have been done on superclusters of similar sizes or larger with open boundary conditions. It was found that edge magnetism\cite{feldner_magnetism_2010} dominates over impurity magnetism. In the periodic scheme, a smaller supercluster implies a higher concentration of impurities. A 42-site supercluster made from the 7 clusters at the center of Fig.~\ref{fig:supercluster} was also tested and the resulting magnetic pattern was similar, albeit with stronger antiferromagnetism far from impurity. A supercluster of $114$ sites may seem excessive and time consuming at first, but the need to include more than one cluster is highlighted by our results. The size of the Hilbert space is exponential ($4^n$ for one cluster) but the time spent on the exact diagonalization is linear in the number of clusters and is easily parallelized. With the geometry considered in Fig.~\ref{fig:supercluster}, the computational bottleneck lies in the matrix inversions \eqref{eq:G} needed for each term in the wavevector sums implied in Eqs~\eqref{latticeAvg} and \eqref{eq:averages}. 
The cost of this inversion is polynomial with complexity $O(N_c^3)$ but the calculation can also be parallelized between different wavevectors.



While one of the main results of this work is that mean field arguments qualitatively capture the correct magnetic correlations induced by non-magnetic impurities in graphene, a second important result is the relation between the on-site interaction $U$ and the threshold values $\epsilon_{0,c}$ and $\epsilon_{0,v}$. This relation could be used to design an experiment to measure how $U$ changes in graphene when screening is modified by different substrates. Indeed, both values $\epsilon_{0,c}$ and $\epsilon_{0,v}$ define a very sharp transition region between the paramagnetic phase and the vacancy-type AFM phase where $\langle \hat{S} \rangle = \frac12$. Since this relation is monotonic, determining either $\epsilon_{0,c}$ or $\epsilon_{0,v}$ is sufficient to obtain an estimate of $U$. Since the critical value $U_c$ is renormalized by long-wavelength fluctuations, our values of $U$ are too small, but we expect that the dependence of $U$ on impurity potential should still be monotonic and linear, as in the inset of Fig.~\ref{fig:AFM_nc}.

In summary, we developed the I-CDMFT technique and applied it to the problem of a nonmagnetic impurity in graphene. This method takes into account electron-electron interactions (dynamical correlations) exactly within small clusters. Our results are consistent with the $T$-matrix treatment and previous mean-field calculations. They have all the key signatures: (1) the net spin $\frac12$ at strong impurity potential imposed by Lieb's theorem, (2) bound states in the LDOS and (3) the correct antiferromagnetic pattern around the impurity. Our work could be straightforwardly extended to study the effect of doping and its influence on Kondo-like screening. This study not only supports previous results on this problem but also benchmarks our new I-CDMFT technique for studying impurity problems while correctly taking into account short-range dynamical correlations. In the future, I-CDMFT could be used to study the effect of impurities in systems where these dynamical correlations produce more important effects, such as high-temperature superconductors, Mott insulators or spin-liquids.

\acknowledgments
We are indebted to E. J. G. Santos for suggesting this problem, to H. Alloul, H. Bouchiat, S. Verret and A. Reymbaut for useful discussions, and to S. Allen for technical help. A.-M.S.T and M.C. are grateful to the Harvard Physics Department where this work was initiated. Work at Sherbrooke was partially supported by NSERC, by a Vanier scholarship from NSERC (M.C.), CIFAR, and the Tier I Canada Research Chair Program (A.-M.S.T.). Partial support by the MIT-Harvard Center for Ultracold Atoms is also acknowledged. Computing resources were provided by CFI, MELS, Calcul Qu\'ebec and Compute Canada. 

%


\end{document}